\documentclass[aps,prmaterials,twocolumn,nofootinbib,floatfix,showkeys]{revtex4-2}

\usepackage{amsmath,amssymb,bm}
\usepackage{graphicx}
\graphicspath{{./}{Documentation/}}
\usepackage{booktabs}
\usepackage{xcolor}
\usepackage{hyperref}

\begin{document}

\title{SR-CGCNN: Shared Recurrent Convolution in Crystal Graph Neural Networks for Materials Property Prediction}

\author{Satadeep Bhattacharjee}
\email{s.bhattacharjee@ikst.res.in}
\affiliation{Indo-Korea Science and Technology Center (IKST), Bengaluru 560065, India}
\date{\today}

\begin{abstract}
Crystal graph neural networks predict materials properties by propagating
information through local atomic environments. In conventional crystal graph
convolutional neural networks (CGCNNs), this propagation depth is increased by
stacking independently parameterized convolutional layers. This coupling
between message-passing depth and parameter count raises a simple question:
can repeated application of the same learned local update recover most of the
benefit of a deeper CGCNN? We address this question by introducing a
shared-recurrent CGCNN (SR-CGCNN), in which the main crystal-graph
convolutional weights are tied across recurrent message-passing steps. The
graph construction, pooling operation, and prediction head are kept unchanged,
allowing a controlled comparison with standard CGCNN baselines. On
Materials Project-derived formation energy and band-gap datasets, a three-step
SR-CGCNN approaches the accuracy of a standard three-layer CGCNN while using
only $34.5\%$ of its trainable convolutional parameters. The formation energy
test mean absolute error changes from $0.0945$ to
$0.0986~\mathrm{eV\,atom^{-1}}$, while the band-gap error changes from
$0.4346$ to $0.4503~\mathrm{eV}$. These results indicate that repeated shared
message passing can provide a parameter-efficient approximation to stacked
CGCNN depth, offering a compact recurrent interpretation of crystal graph
convolution.
\end{abstract}
\maketitle

\section{Introduction}

Crystal structures are naturally represented as graphs: atoms define the nodes, and interatomic neighbor relations define the edges. This representation makes graph neural networks especially well suited to materials-property prediction, and such models have become central tools for learning structure--property relationships directly from atomic geometry~\cite{reiser2022graph,fung2021benchmarking}. In the crystal graph convolutional neural network (CGCNN)~\cite{xie2018crystal}, a crystal is encoded as a graph in which each atom carries an initial feature vector and each neighboring atom pair carries a bond-distance representation. Repeated graph-convolution operations propagate information through the local crystal environment, after which a pooling operation produces a crystal-level embedding for predicting scalar targets such as formation energy, band gap, elastic modulus, or magnetic moment~\cite{bartel2020critical,xu2024machine,park2020developing}. CGCNN established a simple and influential baseline for crystal-property learning because it showed that useful materials representations can be learned directly from structure without hand-crafted descriptors.

For materials applications, the depth of a graph neural network has a direct
interpretation in terms of the spatial extent over which local chemical and
structural information is propagated. A single message-passing step mainly
updates an atomic representation using its near-neighbor coordination
environment, whereas repeated message passing allows information from more
distant coordination shells to influence the final crystal representation.
This is important for properties such as formation energy and band gap, which
depend not only on local composition but also on bonding topology, coordination
geometry, and medium-range structural connectivity. Thus, changing the
message-passing depth is not merely a numerical architectural choice; it also
changes the length scale over which the model can integrate crystal-chemical
information.

In conventional stacked CGCNN models, however, the spatial propagation depth
and the number of independent trainable transformations increase together.
This coupling makes it difficult to determine whether improved prediction
accuracy arises from a larger effective structural receptive field or simply
from a larger number of independent parameters. The distinction is especially
relevant in materials informatics, where many target properties are available
only for moderate-sized datasets and where compact models are desirable for
robust comparison across chemistries. Related efforts have also sought to improve data efficiency and
interpretability in crystalline-materials property prediction through
pretrained crystal representations, autoencoder-based models, and feature
selection, as exemplified by CrysXPP~\cite{das2022crysxpp}. Complementary NLP-based materials-informatics tools such as MatSciRE extract entity--relation triplets from materials-science literature to support
knowledge-base construction~\cite{mullick2024matscire}.

Around the same period and subsequently, several graph-based atomistic architectures were developed that expanded the expressivity of these models by enriching how local environments are represented. SchNet introduced continuous-filter convolutions for atomistic systems, enabling the interaction kernel to depend smoothly on interatomic distances rather than on a fixed discrete edge encoding~\cite{schutt2017schnet}. MEGNet generalized graph-network updates to molecules and crystals and incorporated global state variables, thereby allowing atom, bond, and system-level information to be updated jointly~\cite{chen2019megnet}. ALIGNN further incorporated angular information by coupling the bond graph to its line graph, so that both pairwise and bond-angle relationships could participate explicitly in message passing~\cite{choudhary2021alignn}. Together, these developments show that more expressive message-passing architectures can improve materials modeling. At the same time, they also highlight a complementary question: beyond designing richer local descriptors, how much of the performance gain in stacked graph models comes simply from applying message passing more times?

This question is particularly relevant for the standard stacked-message-passing view of depth. In conventional CGCNN-style architectures, increasing the number of convolutional layers usually increases two things simultaneously: the receptive field over which information can propagate and the number of independent trainable transformations used to process that information. These effects are often entangled. A deeper network may perform better because it accesses a larger structural neighborhood, because it has more independent parameters, or because both changes occur at once. For materials datasets of moderate size, this coupling also affects model efficiency and potentially generalization. The present work is motivated by separating these two roles as cleanly as possible: we seek to preserve message-passing depth while reducing the number of independent convolutional parameter sets.

There is also a natural physical motivation for such a separation. Many computational models in materials science are built around repeated application of a local update rule, as in self-consistent field cycles, iterative structural relaxation, local-field corrections, and charge-equilibration procedures. A graph-neural-network update is not physically equivalent to any of these operations, and we do not interpret it as a surrogate for a specific electronic-structure algorithm. Nevertheless, a recurrent graph convolution is analogous in spirit to these iterative procedures: the same local transformation is applied repeatedly so that information can propagate across progressively larger neighborhoods and the representation can be refined step by step. This viewpoint suggests that a shared update rule may provide a useful inductive bias for crystal graphs, especially when one seeks compact models rather than ever larger stacks of independently parameterized layers.

At the same time, recurrent and weight-tied graph neural networks are not new ideas in graph machine learning. Early graph neural network formulations already treated node representations as the outcome of recurrent local updates~\cite{scarselli2009gnn}, and later recurrent message-passing models likewise emphasized repeated application of a shared relational update rule across multiple reasoning steps~\cite{palm2018rrn}. More recent work has also revisited recurrent graph architectures and analyzed their connections to logical and dynamical perspectives on message passing~\cite{pflueger2024recurrent,li2021dynamics}. Our contribution is therefore not a recurrent GNN in general; rather, it is a recurrent CGCNN for crystal graphs together with a controlled parameter-efficiency comparison against the standard stacked CGCNN architecture.

In this work, we introduce SR-CGCNN, a weight-tied recurrent-depth variant of CGCNN in which a single crystal-graph convolutional block is reused across multiple message-passing steps. We compare CGCNN-1, CGCNN-3, CGCNN-5,SR-CGCNN-3, and SR-CGCNN-5 on Materials Project data~\cite{jain2013commentary} for formation energy and band-gap prediction. The central question is whether shared recurrent convolution can retain most of the accuracy of the standard three-layer CGCNN while using substantially fewer independent convolutional parameters. As we show below, the key result is that SR-CGCNN-3 remains close in accuracy to CGCNN-3 while using only 34.5\% of its convolutional parameters. The architectural distinction between the standard and recurrent models is summarized schematically in Fig.~\ref{fig:schematic}.

\section{Standard Crystal Graph Convolution}

In CGCNN, a crystal structure is represented as a graph
\begin{equation}
    \mathcal{G} = (\mathcal{V}, \mathcal{E}),
\end{equation}
where $\mathcal{V}$ is the set of atoms and $\mathcal{E}$ is the set of neighbor connections. Each atom $i$ is associated with an initial feature vector $\bm{v}_i$, and each neighbor pair $(i,j)$ is associated with an edge feature vector $\bm{u}_{ij}$, typically obtained by expanding the interatomic distance in a Gaussian basis.

The initial atom features are mapped into a hidden embedding space:
\begin{equation}
    \bm{h}_i^{(0)} = W_{\mathrm{emb}} \bm{v}_i + \bm{b}_{\mathrm{emb}} .
\end{equation}
Here, $W_{\mathrm{emb}}\in\mathbb{R}^{w\times d_{\mathrm{in}}}$ and $\bm{b}_{\mathrm{emb}}\in\mathbb{R}^{w}$ are the trainable atom-embedding matrix and bias, respectively, $d_{\mathrm{in}}$ is the dimension of the initial atomic descriptor, and $w$ is the hidden atom-feature dimension.

A crystal graph convolution then updates each atomic embedding by combining the central atom feature, neighbor atom features, and bond features. A typical CGCNN update can be written as
\begin{equation}
    \bm{z}_{ij}^{(t)}
    =
    \bm{h}_i^{(t)} \oplus \bm{h}_j^{(t)} \oplus \bm{u}_{ij},
\end{equation}
where $\oplus$ denotes concatenation. The concatenated feature is passed through a gated transformation:
\begin{equation}
    \bm{m}_{ij}^{(t)}
    =
    \sigma \left( W_f \bm{z}_{ij}^{(t)} + \bm{b}_f \right)
    \odot
    g \left( W_s \bm{z}_{ij}^{(t)} + \bm{b}_s \right),
\end{equation}
Here, $W_f$ and $W_s$ are trainable weight matrices for the gating and message channels, respectively, while $\bm{b}_f$ and $\bm{b}_s$ are their corresponding bias vectors.

where $\sigma$ is a sigmoid gate, $g$ is a nonlinear activation function such as Softplus, and $\odot$ denotes element-wise multiplication. The neighbor messages are summed:
\begin{equation}
    \bm{m}_{i}^{(t)}
    =
    \sum_{j \in \mathcal{N}(i)}
    \bm{m}_{ij}^{(t)} .
\end{equation}
The atomic embedding is then updated with a residual-like transformation:
\begin{equation}
    \bm{h}_i^{(t+1)}
    =
    g \left(
    \bm{h}_i^{(t)} + \bm{m}_i^{(t)}
    \right).
\end{equation}

After $L$ convolutional layers, the atomic embeddings are pooled to form a crystal-level representation:
\begin{equation}
    \bm{h}_{\mathrm{crys}}
    =
    \frac{1}{N}
    \sum_{i=1}^{N}
    \bm{h}_i^{(L)} .
\end{equation}
The pooled representation is passed through fully connected layers to predict the target property:
\begin{equation}
    \hat{y}
    =
    f_{\mathrm{MLP}}
    \left(
    \bm{h}_{\mathrm{crys}}
    \right).
\end{equation}

\section{Shared Recurrent Crystal Graph Convolution}

The proposed modification concerns only the convolutional part of the model. The graph construction, atom features, Gaussian distance expansion, pooling, and final prediction head are kept unchanged. The standard CGCNN uses $L$ independent convolutional blocks:
\begin{equation}
    \bm{h}^{(t+1)}
    =
    \mathrm{Conv}_{\theta_t}
    \left(
    \bm{h}^{(t)}, \mathcal{E}
    \right),
    \qquad
    t=0,\ldots,L-1.
\end{equation}
In contrast, SR-CGCNN uses a single shared convolutional block:
\begin{equation}
    \bm{h}^{(t+1)}
    =
    \mathrm{Conv}_{\theta}
    \left(
    \bm{h}^{(t)}, \mathcal{E}
    \right),
    \qquad
    t=0,\ldots,T-1.
\end{equation}
Here, $T$ is the number of recurrent applications. For the most direct comparison to the standard three-layer CGCNN, we choose $T=3$.

\begin{figure}[t]
    \centering
    \includegraphics[width=\linewidth,height=0.52\textheight,keepaspectratio]{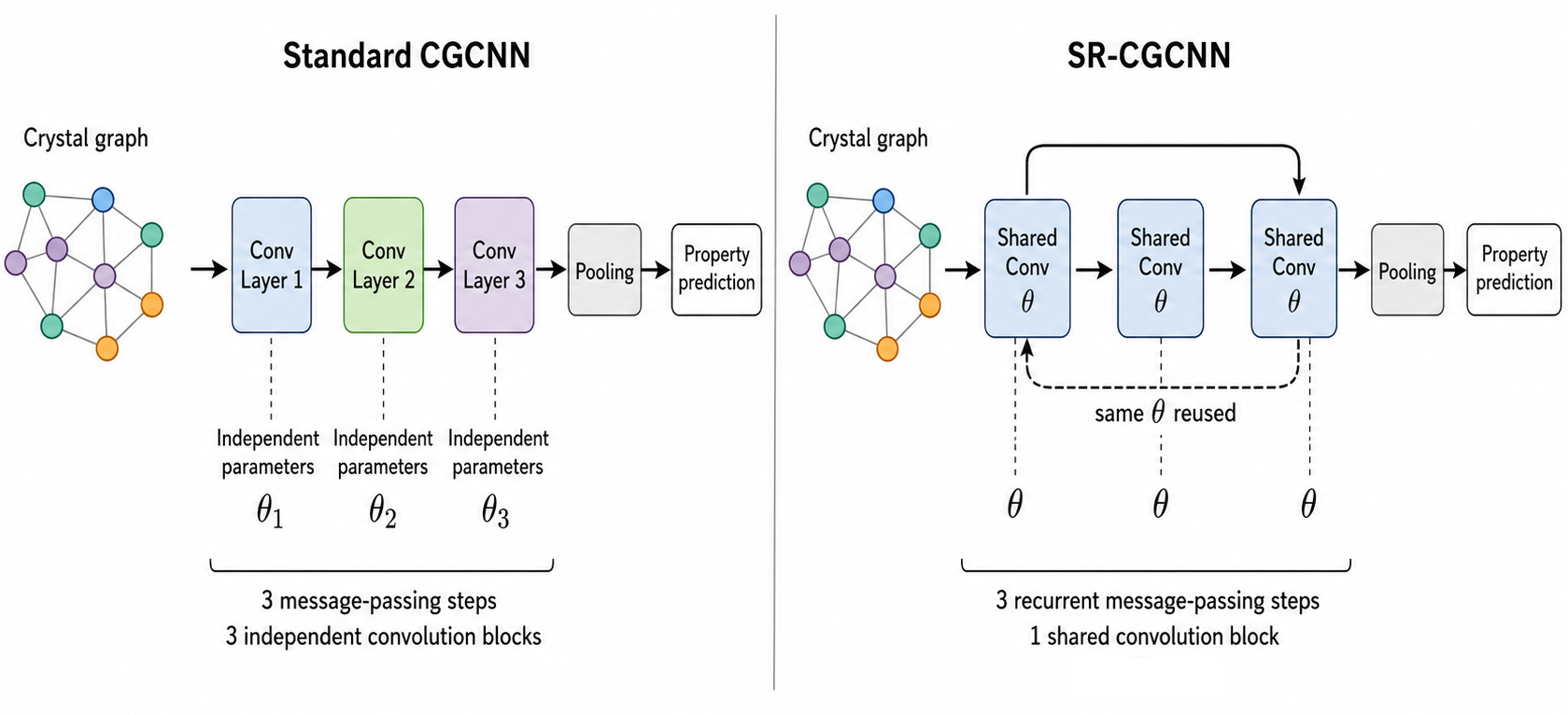}
    \caption{
Schematic comparison between the standard three-layer CGCNN and the proposed
shared-recurrent CGCNN (SR-CGCNN). In the standard architecture, three
message-passing steps are performed using independently parameterized
convolutional blocks with parameters $\theta_1$, $\theta_2$, and $\theta_3$.
In SR-CGCNN, the same message-passing transformation with shared weights
$\theta$ is applied recurrently for the same number of propagation steps.
The graph construction, pooling operation, and prediction head are kept
unchanged. In the implementation used in this work, the main convolutional
weights are shared across recurrent steps, while normalization parameters are
allowed to be step-specific.
}
    \label{fig:schematic}
\end{figure}

For the three-step comparison used in the main benchmarks, the two propagation patterns can be written as
\begin{align}
    \bm{h}^{(3)}_{\mathrm{CGCNN\text{-}3}}
    &=
    \mathrm{Conv}_{\theta_3}
    \circ \mathrm{Conv}_{\theta_2}
    \circ \mathrm{Conv}_{\theta_1}
    \left(\bm{h}^{(0)}\right), \\
    \bm{h}^{(3)}_{\mathrm{SR\text{-}CGCNN\text{-}3}}
    &=
    \mathrm{Conv}_{\theta}
    \circ \mathrm{Conv}_{\theta}
    \circ \mathrm{Conv}_{\theta}
    \left(\bm{h}^{(0)}\right).
\end{align}
Here $\bm{h}^{(t)}$ denotes the set of atomic embeddings after the $t$-th graph-convolution step. The distinction between the two models is shown schematically in Fig.~\ref{fig:schematic}. In the standard model, increasing the number of convolutional layers increases both the message-passing depth and the number of independent parameter sets. In the recurrent model, increasing $T$ increases the effective message-passing depth without increasing the number of independent convolutional parameter sets.

The simplest recurrent update is therefore
\begin{equation}
    \bm{h}_i^{(t+1)}
    =
    \mathrm{CGConv}_{\theta}
    \left(
    \bm{h}_i^{(t)},
    \{\bm{h}_j^{(t)},\bm{u}_{ij}\}_{j\in\mathcal{N}(i)}
    \right).
\end{equation}
This shared-weight update is consistent with the broader recurrent and tied-weight message-passing viewpoint in graph learning~\cite{scarselli2009gnn,palm2018rrn,pflueger2024recurrent}, but here it is specialized to the CGCNN operator for crystal-property prediction.
Since the original CGCNN convolution already contains an internal residual-like update, we do not introduce an additional outer residual connection in the minimal model. This choice keeps the comparison with the standard CGCNN clean.

The recurrent formulation also changes how gradients are accumulated during training. Although SR-CGCNN uses a single convolutional parameter set $\theta$, back-propagation is performed through the unrolled sequence of $T$ graph-convolution steps. Thus the update to $\theta$ receives contributions from every recurrent application:
\begin{equation}
    \frac{\partial \mathcal{L}}{\partial \theta}
    =
    \sum_{t=0}^{T-1}
    \frac{\partial \mathcal{L}}{\partial \bm{h}^{(T)}}
    \left(
    \prod_{s=t+1}^{T-1}
    \frac{\partial \bm{h}^{(s+1)}}{\partial \bm{h}^{(s)}}
    \right)
    \frac{\partial \bm{h}^{(t+1)}}{\partial \theta}.
\end{equation}
This expression emphasizes that weight sharing reduces the number of independent trainable parameters, but it does not remove the intermediate gradient paths associated with the repeated message-passing steps. Each application of the shared CGCNN operator can therefore influence the learned convolutional transformation, while the products of intermediate Jacobians determine how strongly early recurrent steps affect the final loss. As in other recurrent architectures, very large $T$ could in principle lead to attenuated or amplified gradients; in the present work, the main comparison uses the modest depth $T=3$, and the existing CGCNN update and step-specific normalization provide the practical stabilization used in the benchmark models.

Optionally, one may introduce a damped recurrent update:
\begin{equation}
    \bm{h}_i^{(t+1)}
    =
    (1-\alpha_t)\bm{h}_i^{(t)}
    +
    \alpha_t
    \tilde{\bm{h}}_i^{(t+1)},
\end{equation}
where
\begin{equation}
    \tilde{\bm{h}}_i^{(t+1)}
    =
    \mathrm{CGConv}_{\theta}
    \left(
    \bm{h}_i^{(t)},
    \{\bm{h}_j^{(t)},\bm{u}_{ij}\}_{j\in\mathcal{N}(i)}
    \right).
\end{equation}
Here $\alpha_t$ may be fixed or learned. However, this damped version should be treated as a secondary stabilization variant, not as the primary baseline.

\section{Methodology}

\subsection{Model Variants}

The proposed study compares four main model variants:
\begin{table}[t]
\centering
\caption{Model variants for benchmarking.}
\begin{tabular}{lccc}
\toprule
Model & Conv. blocks & Steps & Shared weights \\
\midrule
CGCNN-1 & 1 & 1 & No \\
CGCNN-3 & 3 & 3 & No \\
SR-CGCNN-3 & 1 & 3 & Yes \\
SR-CGCNN-5 & 1 & 5 & Yes \\
\bottomrule
\end{tabular}
\label{tab:model_variants}
\end{table}

The central comparison is between CGCNN-3 and SR-CGCNN-3. Both models perform three message-passing operations, but CGCNN-3 uses three independent convolutional parameter sets, whereas SR-CGCNN-3 uses only one shared convolutional parameter set.

The SR-CGCNN-5 model tests whether increasing recurrent depth beyond the standard three steps leads to additional performance gains or saturation. A shallow CGCNN-1 baseline is included to verify that the recurrent applications are genuinely useful and that the recurrent model is not merely behaving like a one-step graph convolution.

\subsection{Dataset}
The models were evaluated on two structure--property datasets derived from
the Materials Project database~\cite{jain2013commentary}. Crystal structures
and target properties were retrieved from Materials Project and processed using
pymatgen library~\cite{ong2013pymatgen}. The first dataset was used for
formation energy prediction and the second for band-gap prediction. Each data
entry consists of a relaxed crystal structure and a scalar target property:
the formation energy per atom for the formation energy task and the electronic
band gap for the band-gap task.

For both benchmarks, we used 10,000 Materials Project entries. The datasets
should be regarded as fixed task-specific samples from Materials Project rather
than statistically random samples of the full database.  The formation energy targets have
a mean value of $-0.531~\mathrm{eV\,atom^{-1}}$, a sample standard deviation of
$1.030~\mathrm{eV\,atom^{-1}}$, and span the interval
$[-4.508,\,8.993]~\mathrm{eV\,atom^{-1}}$. The band-gap targets have a mean
value of $2.597~\mathrm{eV}$, a sample standard deviation of
$1.628~\mathrm{eV}$, and span the interval $[0.102,\,17.638]~\mathrm{eV}$.

The two target properties were chosen because they probe complementary aspects
of structure--property learning. Formation energy is a thermodynamic quantity
that is strongly influenced by composition, bonding, and structural stability,
whereas the band gap is an electronic property that is sensitive to chemical
identity, orbital hybridization, crystal connectivity, and the underlying electronic-structure
description~\cite{bhattacharjee2024customizing}.

The same graph construction was used for all model variants. Atoms were
represented as graph nodes with elemental feature vectors, near-neighbor
atomic pairs were represented as graph edges, and interatomic distances were
expanded in a Gaussian basis following the original CGCNN formulation. The
crystallographic information files used as CGCNN inputs were generated from
the Materials Project structures using pymatgen. The same
train/validation/test split fractions were used for the formation energy and
band-gap benchmarks, so that differences in performance arise from the model
architecture rather than from changes in data processing.

It should be noted that the purpose of the present
benchmarks is not to establish a full-database leaderboard result, but to
compare the standard and shared-recurrent CGCNN architectures under identical
data and training conditions in a moderate-data regime that remains
computationally accessible and readily reproducible. For the formation energy
benchmark, the retrieved information included the material identifier, relaxed
crystal structure, and formation energy per atom, and no additional
convex-hull stability restriction was imposed. The resulting dataset should
therefore be regarded as a fixed benchmark subset rather than a statistically
representative sample of the complete Materials Project database. The material
identifiers and the corresponding train/validation/test assignments are
provided with the accompanying code to allow the benchmark to be reproduced.

\subsection{Training Protocol}

For each target property, five model variants were trained:
\begin{equation}
\begin{gathered}
    \mathrm{CGCNN\text{-}1}, \quad
    \mathrm{CGCNN\text{-}3}, \quad
    \mathrm{CGCNN\text{-}5}, \\
    \mathrm{SR\text{-}CGCNN\text{-}3}, \quad
    \mathrm{SR\text{-}CGCNN\text{-}5}.
\end{gathered}
\end{equation}
The conventional CGCNN models contain one, three, or five independently
parameterized convolutional blocks, respectively. The SR-CGCNN models reuse
a single main convolutional transformation over three or five recurrent
message-passing steps, while retaining step-specific normalization
parameters.

For each target property, the same material identifiers were assigned to the
training, validation, and test subsets for all model variants. The dataset was
shuffled using the fixed default dataset seed of 123 and subsequently divided
into 60\% training, 20\% validation, and 20\% test subsets.

All models were trained for 300 epochs using the Adam optimizer with a
learning rate of 0.001 and a batch size of 256. The graph construction, atom
and bond features, atom-feature dimension, hidden-feature dimension, pooling
operation, prediction head, and remaining training settings were kept
unchanged across the model variants. The comparisons therefore primarily
probe the effects of message-passing depth and convolutional weight sharing.

\subsection{Evaluation Metrics}

Because both benchmarks are scalar-regression tasks, the primary accuracy
metric is the mean absolute error (MAE), reported in the natural units of each
target property: $\mathrm{eV\,atom^{-1}}$ for formation energy and
$\mathrm{eV}$ for band gap. For a set of $N$ samples with reference values $y_n$ and model
predictions $\hat{y}_n$, the MAE is defined as
\begin{equation}
    \mathrm{MAE}
    =
    \frac{1}{N}
    \sum_{n=1}^{N}
    \left|
    y_n - \hat{y}_n
    \right|.
\end{equation}

MAE is used here because it gives a directly interpretable average prediction
error in physically meaningful units and allows transparent comparison across
the four model variants. Validation MAE is used to monitor model quality during
training and to compare candidate architectures, whereas test MAE is used for
the final benchmark comparison.

However, prediction error alone is not sufficient for the present study. The
central objective is to assess whether recurrent weight sharing can preserve
useful message-passing depth while reducing model complexity. Each model is
therefore evaluated using four quantities: validation MAE, test MAE, the total
number of trainable parameters, and the number of trainable convolutional
parameters. The key comparison is thus an accuracy--parameter trade-off,
especially between the standard three-layer CGCNN and the three-step
SR-CGCNN.

\section{Results and Discussion}

\begin{figure}[t]
    \centering
    \includegraphics[width=0.78\linewidth]{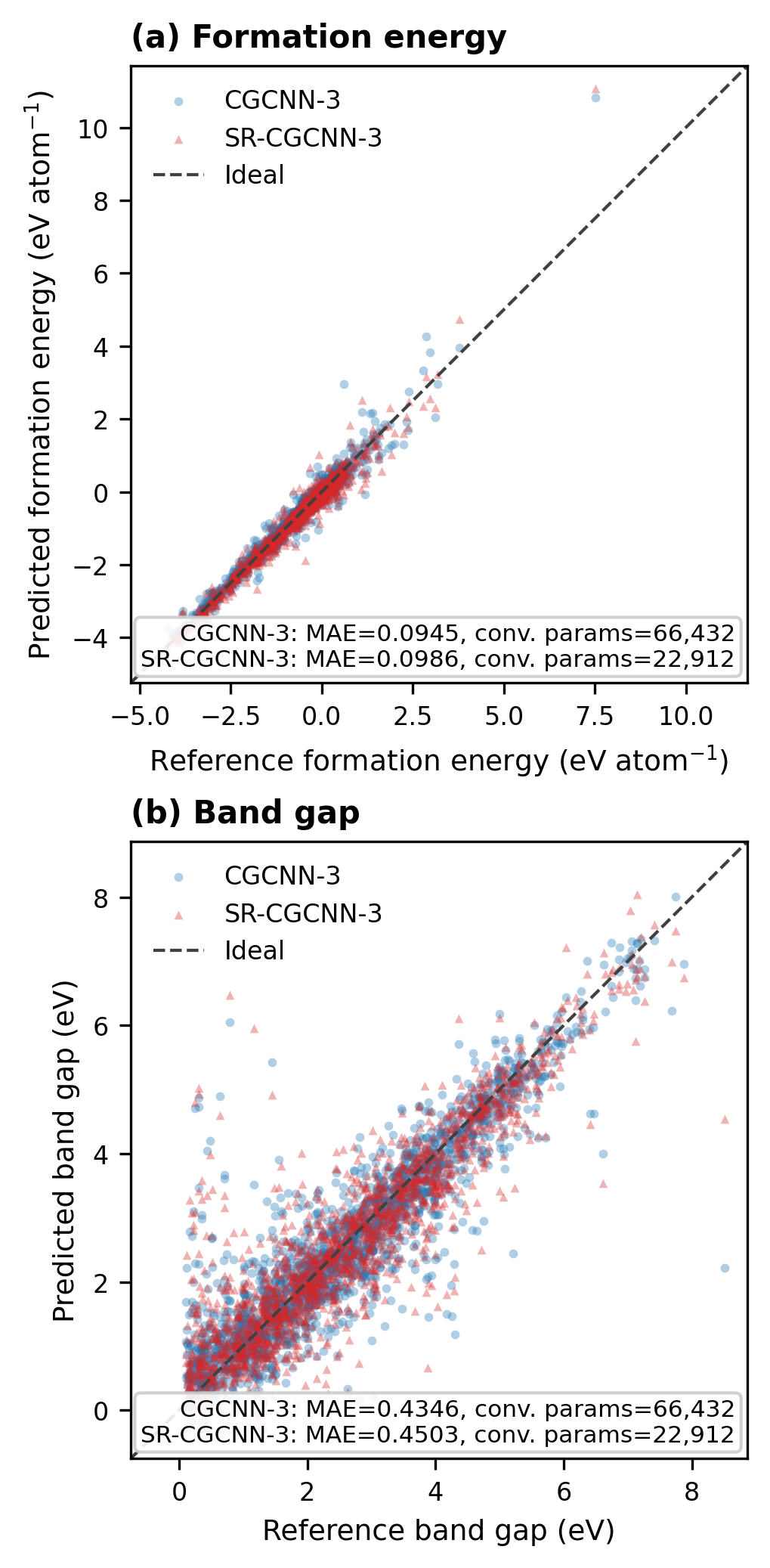}
    \caption{Test-set parity plots for the standard three-layer CGCNN and the
    proposed SR-CGCNN. (a) Formation energy prediction. (b) Band-gap
    prediction. The dashed line denotes ideal agreement between reference and
    predicted values. The three-step SR-CGCNN closely follows the standard
    CGCNN-3 prediction trend for both targets while reducing the number of
    convolutional parameters from 66,432 to 22,912.}
    \label{fig:parity}
\end{figure}

Figure~\ref{fig:parity} provides a compact visual summary of the principal
test-set comparison between the standard CGCNN-3 model and the three-step
SR-CGCNN model for both target properties.

\subsection{Formation Energy Benchmark}

The formation energy calculations were completed for all four model variants
using the same train/validation/test partition and training protocol. The
resulting comparison is summarized in Table~\ref{tab:formation_results}. In
all cases the models were trained for 300 epochs using the Adam optimizer, a
learning rate of 0.001, and a batch size of 256. The data split was 60\% for
training, 20\% for validation, and 20\% for testing.
\begin{table*}[htbp]
\centering
\caption{Formation energy prediction results for standard CGCNN and
SR-CGCNN models. The MAEs are reported in eV atom$^{-1}$.}
\begin{tabular}{lrrrrrr}
\hline
Model & Blocks & Steps & Conv. params & Total params &
Val. MAE & Test MAE \\
\hline
CGCNN-1     & 1        & 1 & 22,144  & 36,545  & 0.1335 & 0.1208 \\
CGCNN-3     & 3        & 3 & 66,432  & 80,833  & 0.1016 & 0.0945 \\
CGCNN-5     & 5        & 5 & 110,720 & 125,121 & 0.1069 & 0.0950 \\
SR-CGCNN-3  & 1 shared & 3 & 22,912  & 37,313  & 0.1055 & 0.0986 \\
SR-CGCNN-5  & 1 shared & 5 & 23,680  & 38,081  & 0.1120 & 0.1031 \\
\hline
\end{tabular}
\label{tab:formation_results}
\end{table*}

The standard three-layer CGCNN gives the lowest test MAE, 0.0945. However,
the three-step shared-recurrent model gives a closely comparable test MAE of
0.0986 while using only 22,912 trainable convolutional parameters. This is
34.5\% of the convolutional parameter count of CGCNN-3, corresponding to a
65.5\% reduction in convolutional parameters. The total trainable parameter
count is also reduced from 80,833 to 37,313. Thus, the three-step recurrent
model retains most of the predictive accuracy of the standard three-layer
model while using less than half of the total number of trainable parameters.

The shallow CGCNN-1 baseline gives a test MAE of 0.1208, which is substantially
worse than both CGCNN-3 and SR-CGCNN-3. This confirms that the recurrent
applications are not merely reproducing a one-step convolution. Instead,
repeated application of the shared convolutional update improves the
representation and recovers much of the benefit normally obtained by stacking
independent convolutional layers. Relative to CGCNN-1, SR-CGCNN-3 reduces the
test MAE by approximately 18.4\%.

The five-step recurrent model gives a test MAE of 0.1031. This remains
competitive with the standard three-layer CGCNN but does not improve over the
three-step recurrent model for the present formation energy task. This
indicates that increasing recurrent depth beyond three steps does not
automatically improve accuracy. The result is consistent with a saturation of
useful message-passing depth, or with the onset of oversmoothing in deeper
recurrent graph updates.

\subsection{Band-Gap Benchmark}

The same four model variants were also evaluated for band-gap prediction. The
training protocol was kept identical to
the formation energy benchmark: 300 epochs, the Adam optimizer, a learning
rate of 0.001, a batch size of 256, and a 60/20/20 train/validation/test
split. The resulting comparison is summarized in
Table~\ref{tab:bandgap_results}.

\begin{table*}[htbp]
\centering
\caption{Band-gap prediction results for standard CGCNN and
SR-CGCNN models. The MAEs are reported in eV.}
\label{tab:bandgap_results}
\begin{tabular}{lrrrrrr}
\hline
Model & Blocks & Steps & Conv. params & Total params &
Val. MAE & Test MAE \\
\hline
CGCNN-1     & 1        & 1 & 22,144  & 36,545  & 0.4895 & 0.5065 \\
CGCNN-3     & 3        & 3 & 66,432  & 80,833  & 0.4335 & 0.4346 \\
CGCNN-5     & 5        & 5 & 110,720 & 125,121 & 0.4316 &
0.4345 \\
SR-CGCNN-3  & 1 shared & 3 & 22,912  & 37,313  & 0.4373 & 0.4503 \\
SR-CGCNN-5  & 1 shared & 5 & 23,680  & 38,081  & 0.4444 & 0.4635 \\
\hline
\end{tabular}
\end{table*}

The standard three-layer CGCNN again gives the lowest test MAE, 0.4346.
However, the three-step shared-recurrent model remains close, with a test MAE
of 0.4503 while using only 34.5\% of the convolutional parameters of
CGCNN-3. The absolute increase in test MAE relative to CGCNN-3 is 0.0157,
while the convolutional parameter count is reduced from 66,432 to 22,912.
Thus, the same parameter-efficiency trend observed for formation energy also
appears for band-gap prediction.

The shallow CGCNN-1 model gives a substantially larger test MAE of 0.5065.
The improvement from CGCNN-1 to SR-CGCNN-3 shows that the recurrent
applications add useful message-passing depth for the band-gap task as well.
The five-step recurrent model gives a test MAE of 0.4635, which remains better
than CGCNN-1 but worse than SR-CGCNN-3. As in the formation energy benchmark,
adding recurrent steps beyond three does not improve the result for the
present setup.

The comparison between CGCNN-1 and SR-CGCNN-3 is particularly informative from
a materials perspective. CGCNN-1 updates each atomic representation using only
one local aggregation step, while SR-CGCNN-3 allows information to propagate
through three successive coordination environments without introducing three
independent convolutional transformations. The improvement of SR-CGCNN-3 over
CGCNN-1 therefore indicates that the additional recurrent steps capture useful
medium-range structural information rather than merely increasing the number
of trainable parameters.

Parameter efficiency is also relevant for practical materials modeling. Many
materials datasets are much smaller and less uniformly distributed than common
machine-learning benchmarks, particularly when target properties require
expensive first-principles calculations or experimental measurements. In such
settings, architectures that preserve physically meaningful message-passing
depth while reducing the number of independent parameters may help reduce
over-parameterization and improve transferability across related materials
families. The present work does not establish such transferability directly,
but it provides a simple architectural route for testing this hypothesis in
future studies.
\subsection{Effect of Increasing Message-Passing Depth}

To assess whether increasing the depth of the conventional CGCNN improves
predictive performance, we also considered a five-layer model containing five
independently parameterized convolutional blocks. For formation energy
prediction, CGCNN-5 gives a test MAE of
$0.0950~\mathrm{eV\,atom^{-1}}$, compared with
$0.0945~\mathrm{eV\,atom^{-1}}$ for CGCNN-3. For band-gap prediction,
the corresponding test MAEs are $0.4345$ and $0.4346~\mathrm{eV}$.
Thus, increasing the conventional CGCNN depth from three to five layers
provides no practically meaningful improvement for either target under the
present training protocol.

At the same five-step message-passing depth, SR-CGCNN-5 gives test MAEs of
$0.1031~\mathrm{eV\,atom^{-1}}$ for formation energy and
$0.4635~\mathrm{eV}$ for band-gap prediction. Although these errors are
higher than those of CGCNN-5, SR-CGCNN-5 uses only 23,680 convolutional
parameters, compared with 110,720 for CGCNN-5, corresponding to a 78.6\%
reduction. The total parameter count is similarly reduced from 125,121 to
38,081, corresponding to a 69.6\% reduction. Taken together with the
SR-CGCNN-3 results, these comparisons indicate that three message-passing
steps provide the most favorable balance between predictive accuracy and
parameter efficiency among the architectures examined here.

\subsection{External Validation on JARVIS-DFT}

To examine whether the observed accuracy--parameter trade-off is specific to
the Materials Project data, we performed an additional formation energy
benchmark using a fixed subset of 10,000 JARVIS-DFT structures~\cite{choudhary2020joint}. CGCNN-3 and
SR-CGCNN-3 were trained using the same graph representation, model dimensions,
data fractions, optimizer, learning rate, batch size, and number of epochs as
in the Materials Project calculations.

CGCNN-3 gives a validation MAE of
$0.1098~\mathrm{eV\,atom^{-1}}$ and a test MAE of
$0.1146~\mathrm{eV\,atom^{-1}}$. The corresponding values for SR-CGCNN-3
are $0.1128$ and $0.1149~\mathrm{eV\,atom^{-1}}$, respectively. Thus, the
difference in test MAE between the two architectures is only
$0.0002~\mathrm{eV\,atom^{-1}}$, corresponding to approximately $0.2\%$
relative to CGCNN-3.

At the same time, SR-CGCNN-3 reduces the number of trainable convolutional
parameters from 66,432 to 22,912, a reduction of 65.5\%, and reduces the
total parameter count from 80,833 to 37,313, a reduction of 53.8\%.
Therefore, the JARVIS-DFT results reproduce the principal trend observed on
the Materials Project benchmarks: recurrent weight sharing retains nearly
the same predictive accuracy while substantially reducing the number of
independently trainable parameters.
\begin{table*}[htbp]
\centering
\caption{Formation energy prediction on the 10,000-structure JARVIS-DFT
benchmark. MAEs are reported in eV atom$^{-1}$.}
\label{tab:jarvis_formation_energy}
\begin{tabular}{lrrrr}
\hline
Model & Conv. params & Total params & Val. MAE & Test MAE \\
\hline
CGCNN-3    & 66,432 & 80,833 & 0.1098 & 0.1146 \\
SR-CGCNN-3 & 22,912 & 37,313 & 0.1128 & 0.1149 \\
\hline
\end{tabular}
\end{table*}

\subsection{Interpretation of the Recurrent Architecture}

The formation energy and band-gap results demonstrate that shared recurrent
convolution is a compact alternative to the conventional stacked CGCNN
architecture. The main comparison is between CGCNN-3 and SR-CGCNN-3, because
both models perform three message-passing operations. The difference is that
CGCNN-3 uses three independently parameterized convolutional blocks, whereas
SR-CGCNN-3 applies a single shared convolutional transformation recurrently.
For formation energy, the observed test MAE increases only from 0.0945 to
0.0986; for band gap, it increases from 0.4346 to 0.4503. In both cases, the
convolutional parameter count decreases from 66,432 to 22,912. This result
suggests that, for the properties studied here, much of the accuracy gain of a
deeper CGCNN arises from repeated local message passing rather than from
completely independent convolutional transformations at every depth.

A crucial implementation detail is the treatment of normalization. In the
shared-recurrent model used for the final calculations, the linear
message-generation layer is shared across recurrent steps, but the batch
normalization layers are step-specific. This distinction is important because
the feature distribution after the first, second, and third recurrent updates
need not be identical. Sharing the full convolutional block, including the
batch-normalization statistics, was found to be too restrictive. In contrast,
sharing the main convolutional transformation while allowing each recurrent
step to maintain its own normalization statistics leads to results that are
comparable to the standard CGCNN baseline.

The comparison with CGCNN-1 is also important because CGCNN-1 and
SR-CGCNN-3 have nearly the same total parameter count (36,545 versus 37,313),
yet SR-CGCNN-3 is substantially more accurate. For formation energy, the test
MAE decreases from 0.1208 to 0.0986, corresponding to
$\Delta \mathrm{MAE} = -0.0222~\mathrm{eV\,atom^{-1}}$. For band-gap
prediction, the test MAE decreases from 0.5065 to 0.4503, corresponding to
$\Delta \mathrm{MAE} = -0.0562~\mathrm{eV}$. Therefore, the recurrent model
is not simply benefiting from a compact parameterization; it benefits from
repeated graph updates. This supports the interpretation of shared recurrent
convolution as an iterative refinement of atomic environment embeddings. From
a materials perspective, the same local update rule is repeatedly applied so
that information can propagate through larger neighborhoods of the crystal
graph, analogous in spirit to iterative refinement procedures used in
atomistic modeling, although the neural update should not be identified with
any specific electronic-structure iteration.

The SR-CGCNN-5 result shows that additional recurrent depth is not always
beneficial. Although the five-step model remains parameter efficient and
reasonably close to CGCNN-3, its test MAE is higher than that of SR-CGCNN-3.
This behavior appears in both benchmarks and suggests that the useful depth may
already be reached at three message-passing steps for these datasets. Deeper
recurrent updates may produce diminishing returns through stronger smoothing of
atom embeddings, long-range information bottlenecks, or related optimization
difficulties, consistent with the oversmoothing and oversquashing literature on
deep graph neural networks~\cite{oono2020optimization,alon2020bottleneck}.
A systematic sweep over recurrent
depth would be useful in future work, but the present calculations already
show that the three-step recurrent model provides the best balance between
accuracy and parameter efficiency among the tested recurrent variants.

Overall, the results support the central hypothesis of this work: an
SR-CGCNN can approach the accuracy of a standard three-layer
CGCNN for materials-property prediction while using substantially fewer
trainable convolutional parameters. The standard CGCNN remains the most
accurate model in absolute terms for both targets, but the recurrent
architecture offers a more compact model with only a small loss in test
accuracy. This trade-off is especially attractive for materials-science
applications where datasets are moderate in size and where compact,
interpretable graph architectures are preferred.

Extending recurrent weight sharing to more elaborate materials architectures
would, however, require model-specific choices. For example, ALIGNN alternates
message passing between the atomistic bond graph and its line graph, the latter
explicitly encoding bond-angle relationships; a recurrent ALIGNN formulation
would therefore require deciding whether the bond-graph and line-graph
transformations should be shared jointly or independently across
depth~\cite{choudhary2021alignn}. Similarly, architectures based on different
structural representations and attention mechanisms would require a distinct
definition of recurrent parameter sharing rather than a direct replacement of
the CGCNN convolutional block. POAT, for example, instead combines pointwise distance-distribution encoding
with an offset-attention mechanism~\cite{yang2025poat}, illustrating that
recurrent parameter sharing outside CGCNN would require an
architecture-specific formulation.

\section{Conclusion}

In this work, we introduced SR-CGCNN, a shared-recurrent variant of CGCNN in
which a single crystal-graph convolutional transformation is reused across
multiple message-passing steps. This design cleanly separates propagation
depth from the number of independent convolutional parameter sets, enabling a
controlled test of whether repeated local updates can recover most of the
benefit of a deeper stacked crystal graph network. Across both formation energy
and band-gap prediction, the main result is consistent: a three-step SR-CGCNN
remains close to the standard three-layer CGCNN in predictive accuracy while
using only 34.5\% of its convolutional parameters. For formation energy, the
test MAE changes from 0.0945 to
$0.0986~\mathrm{eV\,atom^{-1}}$; for band gap, it changes from 0.4346 to
$0.4503~\mathrm{eV}$. At the same time, the comparison with CGCNN-1 shows that
the recurrent architecture is not simply a compact reparameterization: despite
similar total parameter counts, SR-CGCNN-3 achieves lower test error by
benefiting from repeated message passing.

These findings suggest that, for the materials-property tasks considered here,
much of the value of deeper crystal graph models arises from iterative local
information propagation rather than from learning completely independent
convolutional transformations at every depth. The results therefore support
shared recurrent convolution as a simple, parameter-efficient, and physically
interpretable alternative to conventional stacked CGCNN architectures,
especially for moderate-sized materials datasets where compact models are
desirable. At the same time, the weaker performance of SR-CGCNN-5 indicates
that increased recurrent depth is not automatically beneficial and may lead to
diminishing returns. Future work could examine broader depth sweeps, other
materials targets, transfer across chemically diverse datasets, and more
general recurrent crystal graph architectures with learned damping or adaptive
stopping mechanisms.

\appendix

\section{Parameter comparison between CGCNN-3 and SR-CGCNN-3}
\label{app:paramcomparison}

The purpose of this Appendix is to show explicitly how the convolutional and
total trainable parameter counts reported in the main text were obtained. This
comparison is central to the interpretation of the SR-CGCNN architecture,
because CGCNN-3 and SR-CGCNN-3 perform the same number of message-passing
steps but differ in how many independent convolutional parameter sets are
learned.

The comparison is summarized in Table~\ref{tab:appendix_param_comparison}.
Both models use the standard hidden atom-feature width $w=64$ and perform
three message-passing steps. CGCNN-3 uses three independent convolutional
blocks. SR-CGCNN-3 uses one shared main convolutional transformation across
the three recurrent applications, with step-specific normalization parameters.

\begin{table*}[t]
\centering
\caption{Parameter comparison between CGCNN-3 and SR-CGCNN-3.}
\begin{ruledtabular}
\begin{tabular}{lcc}
Model & CGCNN-3 & SR-CGCNN-3 \\
\hline
Message-passing steps & 3 & 3 \\
Independent convolutional blocks & 3 & 1 \\
Shared main convolutional weights & No & Yes \\
Hidden atom-feature width & 64 & 64 \\
Convolutional parameters & 66,432 & 22,912 \\
Total trainable parameters & 80,833 & 37,313 \\
Formation energy test MAE & 0.0945 & 0.0986 \\
Band-gap test MAE & 0.4346 & 0.4503 \\
\end{tabular}
\end{ruledtabular}
\label{tab:appendix_param_comparison}
\end{table*}

Let $d_{\mathrm{in}}$ denote the length of the raw atom descriptor, $d_e$ the
length of the expanded bond feature, $w$ the hidden atom-feature width used by
the convolutional layers, and $h$ the hidden dimension after pooling. In the
present calculations,
\begin{equation}
    d_{\mathrm{in}}=92,\qquad d_e=41,\qquad w=64,\qquad h=128.
\end{equation}
The value $d_{\mathrm{in}}=92$ is the length of the atom-initialization vector
read from the CGCNN \texttt{atom\_init.json} file. The initial embedding maps
each raw atom descriptor to the convolutional hidden space:
\begin{equation}
    \bm{h}_i^{(0)}
    =
    W_{\mathrm{emb}}\bm{v}_i+\bm{b}_{\mathrm{emb}},
    \qquad
    W_{\mathrm{emb}}\in\mathbb{R}^{w\times d_{\mathrm{in}}}.
\end{equation}

For a standard CGCNN convolutional block of hidden width $w$, the concatenated
central-atom, neighbor-atom, and bond feature vector has dimension
\begin{equation}
    \bm{z}_{ij}^{(t)}
    =
    \bm{h}_i^{(t)}\oplus\bm{h}_j^{(t)}\oplus\bm{u}_{ij}
    \in \mathbb{R}^{2w+d_e}.
\end{equation}
The gated transformation maps this vector to two $w$-dimensional
channels:
\begin{equation}
    W_{\mathrm{gate}}
    \in
    \mathbb{R}^{2w\times(2w+d_e)}.
\end{equation}
Including the bias, the batch normalization over the $2w$ gated
features, and the batch normalization over the $w$ summed neighbor features,
one standard CGCNN convolutional block contains
\begin{equation}
    P_{\mathrm{block}}(w)
    =
    2w(2w+d_e) + 2w + 4w + 2w
    =
    4w^2 + 2wd_e + 8w
\end{equation}
trainable parameters.

For CGCNN-3, $w=64$ and there are three independent convolutional blocks.
Therefore,
\begin{align}
    P_{\mathrm{conv}}^{\mathrm{CGCNN\text{-}3}}
    &=
    3P_{\mathrm{block}}(64) \nonumber\\
    &=
    3\left[4(64)^2+2(64)(41)+8(64)\right] \nonumber\\
    &=
    66{,}432.
\end{align}
The total trainable parameter count is
\begin{align}
    P_{\mathrm{total}}^{\mathrm{CGCNN\text{-}3}}
    &=
    (92\times64+64)
    +66{,}432 \nonumber\\
    &\quad
    +(64\times128+128)
    +(128+1) \nonumber\\
    &=
    5{,}952+66{,}432+8{,}320+129 \nonumber\\
    &=
    80{,}833.
\end{align}

For SR-CGCNN-3, the hidden width is also $w=64$, but the main gated 
transformation is shared across the three recurrent applications. The shared
message-generation parameters are
\begin{equation}
    P_{\mathrm{msg}}(64)
    =
    2(64)\left[2(64)+41\right]+2(64)
    =
    21{,}760.
\end{equation}
The batch-normalization parameters are step-specific. At each recurrent step,
the normalization parameters contribute
\begin{equation}
    P_{\mathrm{norm,step}}(64)
    =
    4(64)+2(64)
    =
    384
\end{equation}
trainable parameters. For three recurrent steps,
\begin{align}
    P_{\mathrm{conv}}^{\mathrm{SR\text{-}CGCNN\text{-}3}}
    &=
    P_{\mathrm{msg}}(64)
    +
    3P_{\mathrm{norm,step}}(64) \nonumber\\
    &=
    21{,}760+1{,}152 \nonumber\\
    &=
    22{,}912.
\end{align}
The total trainable parameter count is
\begin{align}
    P_{\mathrm{total}}^{\mathrm{SR\text{-}CGCNN\text{-}3}}
    &=
    (92\times64+64)
    +22{,}912 \nonumber\\
    &\quad
    +(64\times128+128)
    +(128+1) \nonumber\\
    &=
    5{,}952+22{,}912+8{,}320+129 \nonumber\\
    &=
    37{,}313.
\end{align}

Thus, CGCNN-3 and SR-CGCNN-3 have the same hidden width and the same
message-passing depth, but they differ in how convolutional parameters scale
with depth. CGCNN-3 learns three independent convolutional blocks, whereas
SR-CGCNN-3 shares the main message-generation weights across recurrent steps
and keeps separate normalization parameters for each step. This reduces the
trainable convolutional parameter count from $66{,}432$ to $22{,}912$, or to
$34.5\%$ of the CGCNN-3 convolutional parameter count.

\section*{Code Availability}

The codes for this work are available at
\url{https://github.com/sata-deep/SR-CGCNN}.

\bibliography{SR}
\bibliographystyle{apsrev4-2}

\end{document}